\documentclass[preprint,10pt]{sigplanconf}

\usepackage{amsmath}
\usepackage{epsfig}

\begin{document}

\setlength{\pdfpageheight}{\paperheight}
\setlength{\pdfpagewidth}{\paperwidth}

\conferenceinfo{Scala '16}{October 30--31, 2016, Amsterdam, The Netherlands}
\copyrightyear{2016}
\copyrightdata{978-1-nnnn-nnnn-n/yy/mm}
\copyrightdoi{nnnnnnn.nnnnnnn}


\titlebanner{banner above paper title}        
\preprintfooter{short description of paper}   

\title{Monadic Remote Invocation}

\authorinfo{Raphael Jolly}
           {Databeans}
           {raphael.jolly@free.fr}

\maketitle

\begin{abstract}
In order to achieve Separation of Concerns in the domain of remote method
invocation, a small functional adapter is added atop Java RMI, eliminating
the need for every remote object to implement \texttt{java.rmi.Remote} and
making it possible to remotely access existing code, unchanged. The
\texttt{Remote} monad is introduced, and its implementation and usage are
detailed. Reusing the existing, proven technology of RMI allows not to re-invent
the underlying network protocol. As a result, \textit{orthogonal remote
invocation} is achieved with little or no implementation effort.
\end{abstract}




\section{Introduction}

Enterprise application development can be viewed as a struggle to maximize
useful payload code (often referred to as business logic) with respect to an
ever growing, dark mass of boilerplate/generated supporting code, for all the
features expected in the enterprise context (remote access, transactions,
persistence, logging, security, failure recovery, monitoring, etc.) The big
Java enterprise frameworks are devised to automate as much as possible the
handling of these features, making it as transparent as possible. Needless to
say that they only partially succeed in this endeavor, just considering the
enormous complexity of the tools themselves.

One way out of this quagmire might be through Separation of Concerns (SoC)
\cite{Wikipedia:2016} : if one could address one or more of the aforementioned
features separately, it would suddenly become more tractable. This paper is
specifically focused on the topic of remote invocation, which has usually many
adhesions to other concerns.

\section{Outline}

The paper is organized as follows. Section \ref{sec:hist} gives a historical
perspective on RMI. Section \ref{sec:impl} introduces the \texttt{Remote} monad
and its implementation. Section \ref{sec:usage} explains its usage. Section
\ref{sec:mod} details a modification to RMI's generated code that is necessary
in the most common 2-tiered case with non-serializable data. Section
\ref{sec:disc} compares our setting with existing work and Section
\ref{sec:concl} concludes.

\section{State of the art}
\label{sec:hist}

Since its inception, not late after Java itself, RMI \cite{Wollrath:1996} has
been plagued by a several problems, not so hard individually speaking, but which
as a whole resulted in the technology never really being adopted by the
industry. Among these problems are the following:

\begin{enumerate}
\item\label{pbs:java} RMI is Java-only and cannot interact with e.g.
Microsoft technologies, contrary to its former incarnation, CORBA,
which could do so, but at the expense of increased complexity
\item\label{pbs:net} It is not web-friendly and cannot easily
cross proxies and firewalls, due to not using HTTP
\item\label{pbs:dl} An emphasis was made on the code downloading
feature, which incurred security concerns (the same as with applets)
\item\label{pbs:prgm} For the programmer, it is not completely
transparent : the way an object is made remotely accessible is
by extending RMI's \texttt{UnicastRemoteObject}, which
complicates inheritance hierarchies and hinders SoC.
\end{enumerate}

Due to points \ref{pbs:java} to \ref{pbs:dl} above, the industry favored web
services, which are secure, inter-operable with non-Java technology, and can
travel the web (obviously). Meanwhile, JDBC came with its own means of
communication for database access, independent of RMI. As a result, 3-tiered
architectures (JDBC + web front-end) today are still the most prominent model
of enterprise Java applications. The outcome of point \ref{pbs:prgm} is that
remoting code cannot be separated from other concerns in a RMI application,
making the model uncompetitive compared to more straightforward technology
like JDBC, which finally took over.

\begin{table}
\begin{center}
\begin{tabular}{lll}
\hline
DRIVER & ALL JAVA & NETWORK \\
CATEGORY & & CONNECTION \\
\hline
1 - JDBC-ODBC & No & Direct \\
Bridge & &  \\
2 - Native API & No & Direct \\
as basis & &  \\
3 - JDBC-Net & client Yes & Indirect \\
& server Maybe &  \\
4 - Native protocol & Yes & Direct \\
as basis & &  \\
\end{tabular}
\end{center}
\caption{JDBC Driver Categories}
\label{tab:jdbc}
\end{table}

JDBC however does not achieve SoC either, as is clearly visible looking at
the table of driver types (Table \ref{tab:jdbc}), with remote access and SQL
operation always intertwined. One could even argue, that if SQL is still in use
today in the Java world, is because JDBC is so convenient as a network bridge.
Thus, a purely remoting middleware is called for. On the server side, SQL could
be replaced by native/integrated queries, whose implementation would be
simplified, being purely local.

This paper introduces a simple RMI adapter for Scala that follows the Monad
pattern, the \texttt{Remote} monad, aimed at making remote invocations truly
orthogonal, independent from other concerns.

\section{The \texttt{Remote} monad}
\label{sec:impl}

The \texttt{Remote} monad is implemented as a remote object in the RMI sense.
It encapsulates a normal Java object, of type \texttt{T}. A \texttt{Remote}
interface is defined, with type parameter \texttt{T}, extending
\texttt{java.rmi.Remote}, and declaring the two monadic higher-order functions
\texttt{map} and \texttt{flatMap} as remote methods. A third method \texttt{get}
is also provided to ``force'' the monad, that is to bring the encapsulated
value locally.

\begin{verbatim}
@remote trait Remote[+T] extends
    java.rmi.Remote {
  def map[S](f: T => S): Remote[S]
  def flatMap[S](f: T => Remote[S]):
      Remote[S]
  def get: T
}
\end{verbatim}

The remote implementation \texttt{RemoteImpl} extends RMI's
\texttt{UnicastRemoteObject} and implements the above defined
\texttt{Remote} interface.

\begin{verbatim}
class RemoteImpl[T](val get: T) extends
    UnicastRemoteObject with Remote[T] {
  def map[S](f: T => S) = Remote(f(get))
  def flatMap[S](f: T => Remote[S]) = f(get)
}
\end{verbatim}

Remote references of class \texttt{RemoteImpl\_Stub} are obtained through the
\texttt{rmic} utility. In newer Java versions this step can be omitted, as it
is being replaced by the use of a dynamic \texttt{Proxy}.

Remote objects are registered and obtained through factory methods
\texttt{rebind} and \texttt{lookup}, respectively. The \texttt{apply} method
is the monadic ``return'' function.

\begin{verbatim}
object Remote {
  def apply[T](value: T) =
      new RemoteImpl(value)
  def rebind[T](name: String, value: T) =
      Naming.rebind(name, apply(value))
  def lookup[T](name: String) = Naming.
      lookup(name).asInstanceOf[Remote[T]]
}
\end{verbatim}

\section{Usage}
\label{sec:usage}

The basic usage of the \texttt{Remote} monad is as follows. On the server side,
an object is first instantiated and registered through these instructions:

\begin{verbatim}
Remote.rebind("obj", new Object)
println("obj bound in registry")
\end{verbatim}

On the client side, a remote reference to the object is then obtained and
operated like so:

\begin{verbatim}
val ra = Remote.lookup[Object]("obj")
val str = for (a <- ra) yield a.toString
println(str)
// RemoteImpl_Stub[UnicastRef [liveRef:
// [endpoint:[127.0.1.1:49845](remote),
// objID:[7025f768:155b777d4da:-7fee,
// 8868717526554330746]]]]
println(str.get)
// java.lang.Object@494e7ad5
\end{verbatim}

Operations involving more than one object are also possible, as shown below:

\begin{verbatim}
val rb = for (a <- ra) yield new Object
val rc = for (a <- ra; b <- rb) yield
    a.equals(b)
println(rc)
// RemoteImpl_Stub[UnicastRef [liveRef:
// [endpoint:[127.0.1.1:49845](remote),
// objID:[7025f768:155b777d4da:-7fe6,
// 213409570771429818]]]]
println(rc.get)
// false
\end{verbatim}

\begin{figure}
\centering
\epsfig{file=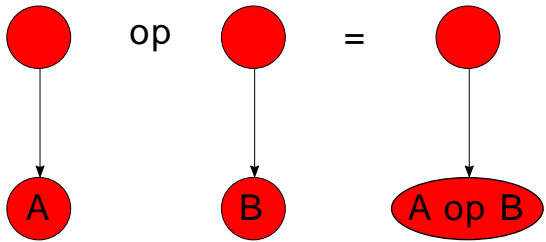,clip=,width=0.85\linewidth}
\caption{Monadic remote invocation}
\label{fig:graph}
\end{figure}

The process is summarized in Figure \ref{fig:graph} : operation \texttt{op} is
sent remotely (query shipping) and applied to objects \texttt{a} and \texttt{b}.
The result is encapsulated in a new remote object and returned for further
operation. Note that encapsulated values need not be \texttt{Serializable}
as long as method \texttt{get} is not used.

\section{Modification of RMI's generated code}
\label{sec:mod}

In order to support the most common case of a 2-tiered application, a small
adjustement to RMI's generated code has to be made. When a remote reference
returns to the place where it was first instantiated, and contrary the
intuition, it is not replaced by its implementation. This specifically causes
a problem when two or more remote objects are operated on, as in the last
example, which is desugared to:

\begin{verbatim}
val rc = ra.flatMap(a => rb.map(b =>
    a.op(b)))
\end{verbatim}

This corresponds to the following sequence of actions:

\begin{itemize}
\item Closure \texttt{a => ...} is sent to remote object \texttt{ra}'s location,
capturing remote object \texttt{rb}.
\item From \texttt{ra}'s location, closure \texttt{b => ...} is addressed to
\texttt{rb} via its method \texttt{map}, capturing local object \texttt{a}
\item Even if \texttt{rb} resides on the same location as \texttt{ra}, method
\texttt{map}'s receiver is a remote reference, not an implementation.
\item Consequently, the call to \texttt{map} induces the serialization
of closure \texttt{b => ...}, and henceforth of object \texttt{a}
\item Object \texttt{a} is deserialized once arrived at \texttt{rb}'s location
\item Operation \texttt{op} is performed on \texttt{a} and \texttt{b}, and the
result is encapsulated and returned.
\end{itemize}

Hence, when \texttt{ra} and \texttt{rb} are in the same location, we have an
opportunity to prevent serialization of the content of \texttt{ra}, if we
replace remote references by their implementation when they come back home.
This is done through the addition of method \texttt{readResolve} below to
\texttt{rmic}'s output \texttt{RemoteImpl\_Stub}.

\begin{verbatim}
public Object readResolve() throws
        java.io.ObjectStreamException {
    return Remote$.MODULE$.replace(this);
}
\end{verbatim}

Method \texttt{replace} above is implemented in \texttt{Remote}'s companion
object using Java's \texttt{WeakReference} mechanism, and method \texttt{apply}
is modified accordingly, storing newly created remote objects in the cache
prior to returning them.

\begin{verbatim}
val cache: Map[java.rmi.Remote,
    Reference[Object]] = new WeakHashMap[
    java.rmi.Remote, Reference[Object]]()

def apply[T](value: T) = {
  val obj = new RemoteImpl(value)
  cache.put(obj, new WeakReference(obj))
  obj
}

def replace[T](obj: Remote[T]): Object = {
  val w = cache.get(obj)
  if (w == null) obj else {
    val o = w.get()
    if (o == null) obj else o
  }
}
\end{verbatim}

It must be noted that since \texttt{Remote} is the only RMI object that will
have to be synthesized in our setting, the manipulation is done once and for
all.

\section{Discussion}
\label{sec:disc}

The \texttt{Remote} monad operates synchronously, which means calls to methods
\texttt{map} and \texttt{flatMap} are blocking. To avoid network delays, it is
relevant to make it asynchronous. It turns out to be one more feature that can
be factored out and implemented separately, namely using the \texttt{Future}
monad \cite{Haller:2012}. By prepending the \texttt{Async} adapter below to
the \texttt{Remote} monad, the resulting model is semantically close to
\texttt{Future}, that is, execution is done in parallel on a different
thread/processor, just also on a different host.

\begin{verbatim}
class Async[T](val value:
    Future[Remote[T]]) {
  def map[S](f: T => S) = Async(value.map {
    remote => remote.map(f)
  })
  def flatMap[S](f: T => Async[S]) =
      Async(value.map {
    remote => remote.flatMap {
      t => Await.result(f(t).value,
          Duration.Inf)
    }
  })
  def get = value.map {
    remote => remote.get
  }
}
\end{verbatim}

Class \texttt{Async} takes a constructor argument of type \texttt{Future}
of \texttt{Remote} and encapsulates it. Its method \texttt{map} forwards the
call to the \texttt{Remote} and re-encapsulates the result. So does method
\texttt{flatMap}, but with a modified function argument that forces (awaits for)
the returned \texttt{Future}. Method \texttt{get} again forwards the call to the
\texttt{Remote}, that is, forces it (brings it locally). Its return value has
type \texttt{Future} and can be used as is for further asynchronous operation,
or forced to get the underlying value. Function \texttt{apply} below is
overloaded to allow instantiation from both a \texttt{Remote} and a
\texttt{Future} of \texttt{Remote}.

\begin{verbatim}
object Async {
  def apply[T](value: Remote[T]): Async[T] =
      apply(Future(value))
  def apply[T](value: Future[Remote[T]]) =
      new Async(value)
}
\end{verbatim}

Likewise, the Function Passing model \cite{Miller:2015} introduces a monadic
remoting concept that is similar to ours, but that in addition features
deferred application as one of its core principles, in order to save time and
space. This feature can also be factored out and separated from purely remoting
concerns, which is done through the \texttt{Deferred} adapter below.

\begin{verbatim}
class Deferred[U, T](val remote: Remote[U],
    val op: U => T) {
  def map[S](f: T => S) = Deferred(remote,
      f.compose(op))
  def flatMap[S](f: T => Deferred[U, S]) =
      f(get)
  def get = remote.map(op).get
}
\end{verbatim}

Class \texttt{Deferred} takes a \texttt{Remote} and a function as constructor
arguments. Value member \texttt{op} is used as an accumulator for functions
passed to method \texttt{map}, which then re-encapsulates it together with
the \texttt{Remote} object. Method \texttt{flatMap} just forces the monad and
applies its function argument \texttt{f} to the result. Method \texttt{get}
applies the accumulated function to the \texttt{Remote} object and forces
it (brings it locally). Method \texttt{apply} is again overloaded, to allow
initial instantiation from a \texttt{Remote}, and re-instantiations from
the same \texttt{Remote} and a function accumulator \texttt{op}.

\begin{verbatim}
object Deferred {
  def apply[T](remote: Remote[T]):
      Deferred[T, T] = apply(remote,
      identity[T])
  def apply[U, T](remote: Remote[U],
    op: U => T) = new Deferred(remote, op)
}
\end{verbatim}

\section{Conclusion}
\label{sec:concl}

We have presented a functional programming equivalent to Java's Remote Method
Invocation in the form of a \texttt{Remote} monad, which allows accessing
existing business logic remotely without any change. By re-using RMI as
the underlying network stack, the implementation was made straightforward
and simple. We have shown how our setting can be easily adapted to reproduce
the operation of similar models like the \texttt{Future} monad or the Function
Passing model.




\bibliographystyle{abbrvnat}
\bibliography{paper}

\end{document}